\documentclass[sigconf]{acmart}

\usepackage{balance}
\usepackage{bm}
\usepackage{threeparttable}
\usepackage{booktabs} 
\usepackage{multirow}
\usepackage{subfigure}
\usepackage{enumitem}
\usepackage{balance}
\usepackage{diagbox}
\usepackage{enumitem}
\setlist[itemize]{leftmargin=*}

\AtBeginDocument{%
  }

\setcopyright{acmlicensed}
\copyrightyear{2018}
\acmYear{2018}
\acmDOI{XXXXXXX.XXXXXXX}

\acmConference[Conference acronym 'XX]{Make sure to enter the correct
  conference title from your rights confirmation emai}{June 03--05,
  2018}{Woodstock, NY}
\acmISBN{978-1-4503-XXXX-X/18/06}

\begin{document}

\title{Coarse-to-fine Dynamic Uplift Modeling for Real-time Video Recommendation}


\author{Chang Meng}
\authornote{Both authors contributed equally to this research.}
\affiliation{
  \institution{Kuaishou Technology}
  \city{Beijing}
  \country{China}
}
\email{mengchang@kuaishou.com}

\author{Chenhao Zhai}
\authornotemark[1]
\authornote{Work done when he was a research intern at Kuaishou Technology.}
\affiliation{
  \institution{Shenzhen International Graduate School, Tsinghua University}
  \city{Shenzhen}
  \country{China}
}
\email{dch23@mails.tsinghua.edu.cn}

\author{Xueliang Wang}
\affiliation{
  \institution{Kuaishou Technology}
  \city{Beijing}
  \country{China}
}
\email{wangxueliang03@kuaishou.com}

\author{Shuchang Liu}
\affiliation{
  \institution{Kuaishou Technology}
  \city{Beijing}
  \country{China}
}
\email{liushuchang@kuaishou.com}

\author{Xiaoqiang Feng}
\affiliation{
  \institution{Kuaishou Technology}
  \city{Beijing}
  \country{China}
}
\email{fengxiaoqiang@kuaishou.com}

\author{Lantao Hu}
\authornote{The corresponding author.}
\affiliation{
  \institution{Kuaishou Technology}
  \city{Beijing}
  \country{China}
}
\email{hulantao@kuaishou.com}

\author{Xiu Li}
\authornotemark[3]
\affiliation{
  \institution{Shenzhen International Graduate School, Tsinghua University}
  \city{Shenzhen}
  \country{China}
}
\email{li.xiu@sz.tsinghua.edu.cn}

\author{Han Li}
\authornotemark[3]
\affiliation{
  \institution{Kuaishou Technology}
  \city{Beijing}
  \country{China}
}
\email{lihan08@kuaishou.com}

\author{Kun Gai}
\affiliation{
  \institution{Unaffiliated}
  \city{Beijing}
  \country{China}
}
\email{gai.kun@qq.com}

\renewcommand{\shortauthors}{Chang Meng et al.}

\begin{abstract}
  With the rise of short video platforms, video recommendation technology faces more complex challenges. Currently, there are multiple non-personalized modules in the video recommendation pipeline that urgently need personalized modeling techniques for improvement. Inspired by the success of uplift modeling in online marketing, we attempt to implement uplift modeling in the video recommendation scenario. 
  However, we face two main challenges: 1) Design and utilization of treatments, and 2) Capture of user real-time interest. To address them, we design \emph{adjusting the distribution of videos with varying durations} as the treatment and propose Coarse-to-fine Dynamic Uplift Modeling (CDUM) for real-time video recommendation. CDUM consists of two modules, CPM and FIC. The former module fully utilizes the offline features of users to model their long-term preferences, while the latter module leverages online real-time contextual features and request-level candidates to model users’ real-time interests. These two modules work together to dynamically identify and targeting specific user groups and applying treatments effectively. Further, we conduct comprehensive experiments on the offline public and industrial datasets and online A/B test, demonstrating the superiority and effectiveness of our proposed CDUM. Our proposed CDUM is eventually fully deployed on the Kuaishou platform, serving hundreds of millions of users every day. The source code will be provided after the paper is accepted.
\end{abstract}





\maketitle
\section{INTRODUCTION}\label{introduction}
The widespread popularity of short videos on social media presents new opportunities and challenges for optimizing recommendation systems on video platforms. During the video recommendation process, users show varying preferences for different types of videos through multi-faceted responses (e.g., watch time and various types of interactions). To ensure a great user experience, a common approach in the industry is to design specific modules within online recommendation pipelines to meet users' complex demands for different videos. However, most of these modules are non-personalized. For example, to provide users with a more diverse range of videos, the current online pipeline employs modules that adjust the distribution of videos with different durations. These modules categorize videos based on their durations and use \emph{uniform parameters} to adjust the number of videos within different categories, which can not meet the varying preferences of different users for the duration of videos. 
To address such issues, we have deployed various strategy-based solutions. However, these approaches struggle to capture long-term user preferences and lack robust generalization, which hampers effective exploration and learning of user interests.

\begin{figure*}[t]
	\centering
	\setlength{\belowcaptionskip}{-0.0cm}
	\setlength{\abovecaptionskip}{-0.0cm}
	\includegraphics[width=0.85\linewidth]{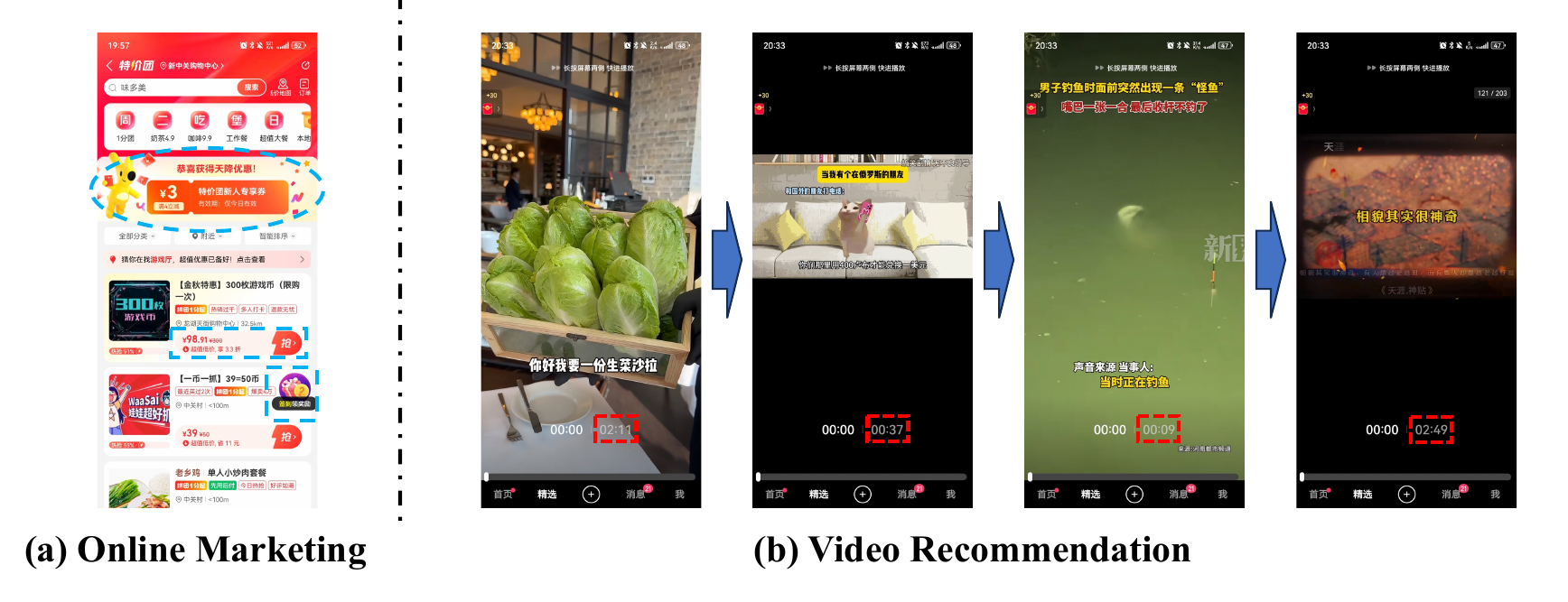}
	\caption{A example of the difference between online marketing and video recommendation scenarios.}
	\label{fig:example}
	\vspace{-3mm}
\end{figure*}

Uplift modeling, as a personalized modeling technique for populations, is used to measure the incremental impact of treatments or interventions for populations \cite{efin,flextenet,dragonnet,euen}. Unlike traditional models, uplift modeling identifies the causal effect of an action by comparing treated and controlled groups, allowing for more targeted decision-making. In industrial scenarios, uplift modeling is widely applied in online marketing scenarios, aiming to provide targeted users with specific incentives, such as coupons, discounts, and bonuses. These methods are based on meta-learning models (such as T-learner \cite{metalearners} and S-learner \cite{metalearners}), tree-based models \cite{zhao2017uplift,radcliffe2011real,athey2016recursive} and Deep Uplift Modeling (DUM) \cite{cevae,efin,descn,euen,liu2024benchmarking,xu2022learning}. They utilize offline user features for uplift modeling to estimate the individual treatment effect (i.e., uplift). By identifying and targeting the users that are most responsive and positively influenced by incentives, these methods significantly reduce marketing costs while improving metrics such as commission conversion rates.


Inspired by the success of uplift modeling in online marketing scenarios, we attempt to apply this approach for video recommendation. Nevertheless, as illustrated in Figure \ref{fig:example}, there are significant differences between video recommendation scenarios and online marketing. In online marketing, the scenario relies solely on users' historical behavior data, with promotional information directly linked to products or services, making treatment and response modeling relatively straightforward. However, in the video recommendation, real-time dynamics are more pronounced. User preferences for the duration and content of the videos shift rapidly, as a variety of videos are pushed to the user (shown on the right side of Figure \ref{fig:example}). Besides, the responses of users are also diverse and complex, thus it becomes more challenging to effectively align and associate treatments with responses. Due to the differences in scenarios, existing uplift modeling techniques cannot be seamlessly integrated or implemented, presenting the following challenges:
\begin{itemize}
    \item Design and utilization of treatments. In the video recommendation scenarios, there are diverse and complex user responses, such as watch time, app usage duration, and video view count. These responses often constrain each other and may even exhibit inverse correlations. Therefore, it's essential to design appropriate treatments to balance these trade-offs among responses. Moreover, the thoughtful design and utilization of treatments can further aid in modeling long-term user preferences.
    \item Capture of user real-time interest. In video recommendation scenarios, modeling both long-term preferences and real-time interests has always been a key focus for us. The former represents the user's historical preference profile, which is more coarse-grained, while the latter captures the user's immediate interest feedback, offering a finer granularity of information. Existing uplift modeling techniques can utilize daily-level features, such as watch time over the past 7 or 14 days, to model a user's long-term preferences. However, these techniques are not capable of capturing real-time user interests effectively. On the other hand, existing methods often have biases when modeling long-term preferences. Designing effective paradigms to capture real-time interests can dynamically correct these biases, thereby enhancing the user's long-term experience.
\end{itemize}

To address these two challenges, we carefully design the treatment based on the specific characteristics of the video recommendation scenario and propose \underline{\textbf{C}}oarse-to-fine \underline{\textbf{D}}ynamic \underline{\textbf{U}}plift \underline{\textbf{M}}odeling for Real-time Video Recommendation (CDUM), which includes two modules: Coarse-grained Preference Modeling (CPM) and Fine-grained Interest Capture (FIC).

To address the first challenge, we recognize that video duration distribution adjustment modules are extensively integrated across various stages of the online pipeline. By increasing the exposure ratio of short videos, it can effectively boosts commercialization metrics, while enhancing the exposure of long videos helps improve consumption metrics. This trade-off underscores the importance of optimizing the exposure ratio of videos of different durations, thereby catering to users' diverse consumption preferences and achieving more precise recommendation outcomes. Therefore, we ultimately decided to use \emph{adjusting the distribution of videos with varying durations} as the treatment. To fully exploit the semantic information of the treatment, we draw inspiration from multi-interest modeling and design the CPM module with a multi-treatment learning paradigm. This module expands the treatment representation into two parts: guidance and indicator. The guidance part is used to filter and extract the information from user features, while the indicator part enhances the model's generalization across different treatments. The CPM module makes full use of users' offline features to model their long-term preferences and outputs these long-term preference scores to the FIC module.

To tackle the second challenge, we design the FIC module, which uses online real-time contextual features and request-level candidates to model users' current interests. It is mainly constructed by a MTL module, and outputs real-time interest scores for videos with different durations to finely adjust the long-term preference scores from the CPM module, dynamically refining the target users for the treatment. The incorporation and utilization of online real-time features not only correct biases in long-term preference modeling but also capture users' real-time information in a fine-grained manner.

In summary, our work makes the following contributions:
\begin{itemize}
    \item We highlight the difficulties and challenges faced in implementing uplift modeling in video recommendation scenarios, specifically regarding the design and utilization of treatments and the capture of users' real-time interests. Additionally, based on current online marketing scenarios, we have proposed insightful ideas for treatment design and utilization in the context of video recommendations.
    \item We have innovatively proposed Coarse-to-fine Dynamic Uplift Modeling for real-time video recommendation. This approach combines offline features with online real-time features to model users' long-term preferences and real-time interests in a coarse-to-fine paradigm, dynamically identifying and targeting specific user groups and applying treatments effectively.
    \item We test our model on two public datasets, an industrial dataset from Kuaishou, and through online A/B experiments on the Kuaishou platform. Our model not only demonstrates superior offline performance but also achieves significant improvements in consumption metrics and user retention on Kuaishou. Our technique is ultimately deployed on the online platform, serving hundreds of millions of users.
\end{itemize}
\section{RELATED WORK}
\label{related_work}
Uplift models aim to identify the target user groups for each specific treatment by accurately estimating the individual treatment effects (ITE). Existing Uplift modeling approaches fall into three categories: meta-learner based methods, tree-based methods, and neural network-based methods.

Meta-learner based methods \cite{yao2021survey,metalearners} are a classic framework that focuses on using existing prediction methods to learn user responses. S-learner \cite{metalearners} and T-learner \cite{metalearners} are two commonly used meta-learning paradigms. The S-learner combines the treatment variable with other features and estimates ITE through a single model. The T-learner builds models separately for the control and treatment groups and estimates ITE through the subtraction between the two models. In both paradigms, when there is a significant imbalance in the amount of data between the treatment and control groups, this imbalance may significantly affect model performance, thereby affecting the estimation of ITE. To address this issue, the X-learner \cite{metalearners} method has been improved based on the T-learner. It introduces the concept of propensity scores and two additional ITE estimators, combining propensity scores to weight the estimation of ITE. This method performs particularly well when there is a significant difference in the number of samples between the treatment and control groups, or when the ITE pattern is relatively simple. Tree-based methods \cite{zhao2017uplift,radcliffe2011real,athey2016recursive,nandy2023generalized} utilize specific tree or forest structures to progressively separate sensitive subgroups corresponding to each treatment from the entire population using different metric-based splitting criteria, with treatment information included in the calculations of the splitting process.

Due to the outstanding feature extraction capabilities of deep learning, methods based on neural networks \cite{wei2024multi,cevae,bica2020estimating,sun2024m,dragonnet} have become a popular framework for uplift estimation. They introduce more complex and flexible architectures to model the response process to treatment, enabling the learning of more accurate user responses or uplift estimation values. These methods are widely applied in industries such as online marketing and precision medicine \cite{he2024rankability,huang2024entire,liu2024benchmarking,zhou2023direct}. For instance, in online marketing, these methods utilize offline user features to perform uplift modeling, and select the target population that is sensitive to incentives (e.g. coupons, discounts, and bonuses) and have a positive effect based on the estimated ITE value (i.e. uplift), which reduces marketing costs while significantly improving indicators such as commission conversion rate. For example, EUEN \cite{euen} uses two sub-neural networks to estimate the conversion probability of the control group and the uplift effect respectively; DESCN \cite{descn} is an end-to-end multi-task cross-network designed to comprehensively capture the relationship between treatment and real response while alleviating treatment bias and sample imbalance issues; EFIN \cite{efin} focuses on explicit feature interaction awareness, especially the sensitivity of non-treatment features to specific treatments, and enhances the model's robustness through an intervention constraint module.

However, all of the above methods are limited to using offline features for uplift modeling and identifying sensitive target population, which means they fail to capture users' real-time interests, thereby being unable to dynamically adjust the target population with specific treatment. Moreover, the design and utilization of treatments in these methods cannot be directly transferred to the context of video recommendations. Our proposed CDUM designs a treatment plan for the video recommendation scenario. It initially models users' long-term preferences at a coarse granularity using daily-level features, then refines the uplift modeling by capturing users' real-time interests through online real-time features, dynamically adjusting the target population for the specific treatment. 
\section{PRELIMINARIES}
Since our designed model consists of both offline and online components, in this section, we will define and explain the problem formulation for both offline and online scenarios separately.
\subsection{Offline Problem Definition}
\label{Offline Problem Definition}
We denote $\mathcal{Z}=\{\mathbf{z}_1, \mathbf{z}_2, ..., \mathbf{z}_N\}$ as the set of instances at the video recommendation scenarios, where $\mathbf{z}_i = (\mathbf{x}_i,\mathbf{t}_i,y_i) \in \mathcal{X} \times \mathcal{T} \times \mathcal{Y}$ and $N$ is the number of training instances. $\mathbf{x}_i \in \mathbb{R}^{f_x \times 1}$ represents the $f_x$-dimensional user features and contextual features of the $i$-th instance. While $\mathbf{t}_i \in \mathbb{R}^{f_t \times 1}$ is the $f_t$-dimensional treatment features for the $i$-th instance. Besides, $y_i \in \{y_i^0,y_i^1,...,y_i^K\}$ is the response label under specific treatment for the $i$-th instance. Besides, we consider the first treatment feature of each instance to represent the treatment's index ID, with a total of $K$ treatments, i.e., $t_i^0 \in [0,K]$.

Following the Neyman-Rubin potential outcome framework \cite{neyman_rubin}, we denote $y_i^k$ and $y_i^0$ as the potential outcome when the user in the $i$-th instance gets a particular treatment $t_i^0 \in [1,K]$ or is not treated, respectively. 
As we can not observe $y_i^k$ and $y_i^0$ at the same time, there is no true uplift result for each instance, which is known as a key reason uplift modeling differs from traditional supervised learning. Thus, we need uplift modeling to estimate the expected individual treatment effect $\tau_k(x_i)$ for each instance.

Different from the existing uplift methods which estimates $\tau_k(x_i)$ with the calculation of the difference between $\mathbb{E}(y_i^k|t_i^0=k,x_i)$ and $\mathbb{E}(y_i^0|t_i^0=k,x_i)$, we pass $\mathbb{E}(y_i^k|t_i^0=k,x_i)$ and $\mathbb{E}(y_i^0|t_i^0=k,x_i)$ into online and combine online module (CPM) to estimate $\tau_k(x_i)$. As a result, we can dynamically refine the estimation of ITE scores and adjust the effective users under different treatments in real-time.

\subsection{Online Problem Definition}
\label{Online Problem Definition}
In the online video recommendation pipeline, we have real-time contextual features and request-level candidates, which are represented as $\mathbf{s}_i=[\mathbf{s}_i^0,\mathbf{s}_i^1,...,\mathbf{s}_i^{f_s}]$. Here $\mathbf{s}_i^j \in \mathbb{R}^{L^j \times 1}$ represents the $j$-th contextual features or request-level candidates for the $i$-th instance, and $L^j$ represents the number of recent videos with which the user has interacted. It is worth noticing that we leverage a request-level instance to train the online module, i.e., modeling an instance using the features and label of each user request.
On this basis, we further design $r_i^k \in \{r_i^1,...r_i^K\}$ as the request-level label of online module. Here $K$ corresponds to the number of treatments. To model the real-time interest of the users dynamically, $r_i^k$ is designed to be calculated from the interaction data of exposed videos within the user candidate set at the request-level. Specifically, we divide the exposed videos into $K$ categories based on their durations, which directly corresponds to the design of our treatment. Next, for the $k$-th duration category, we calculate the ratio of the number of long-play to short-play videos for each user and denote it as $r_i^k$. Additionally, we use $\hat{r}_i^k$ as the real-time interest score, representing the user's preference for videos of different durations. This score is used to adjust the offline long-term preference score 
$\mathbb{E}(y_i^k|t_i^0=k,x_i)$, allowing us to estimate the user's individual treatment effect scores. Finally, we can rank leverage this score to make a rational treatment assignment.

\section{METHOD}
\label{method}

\begin{figure*}[t]
	\centering
	\setlength{\belowcaptionskip}{-0.0cm}
	\setlength{\abovecaptionskip}{-0.0cm}
	\includegraphics[width=0.8\textwidth]{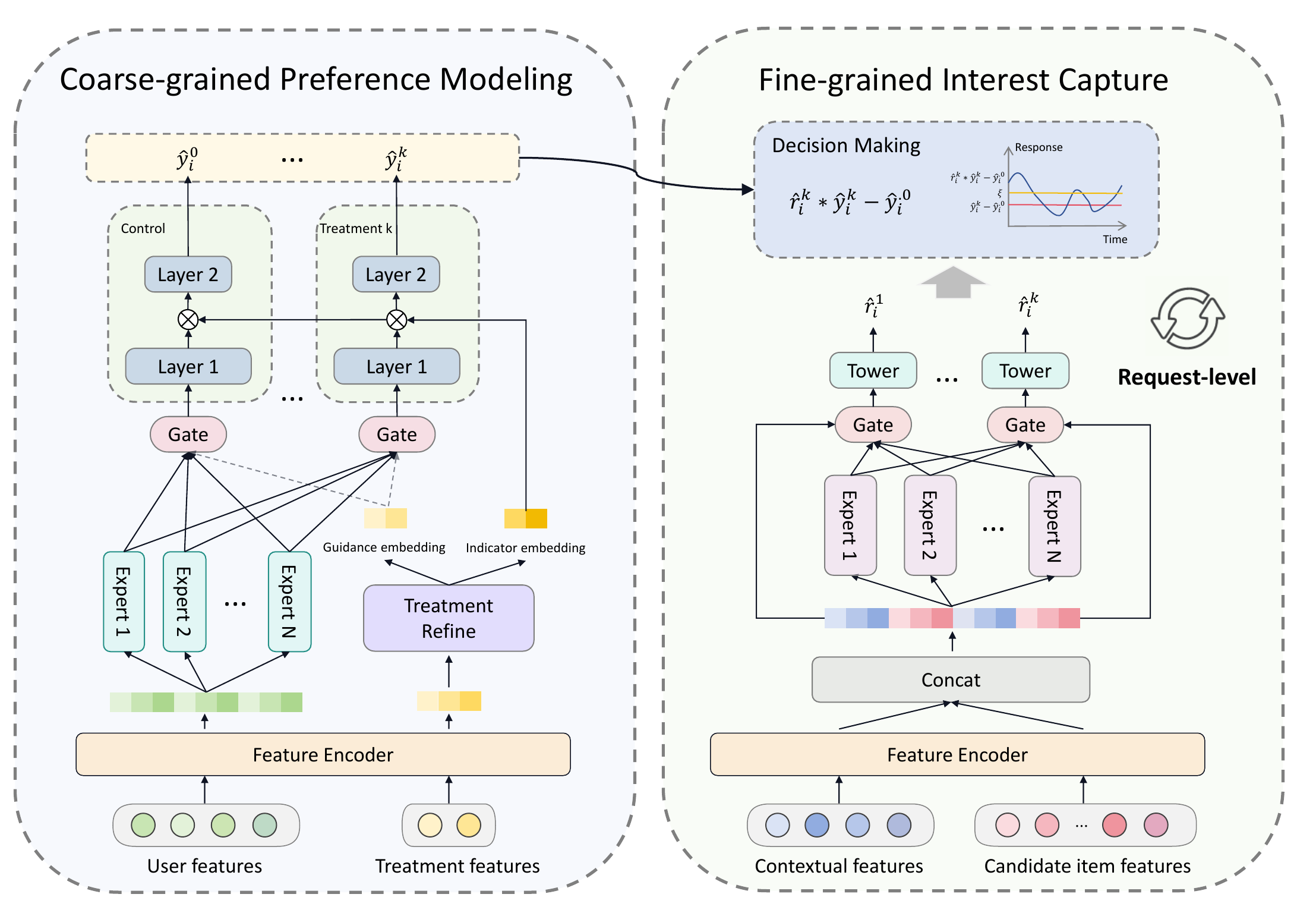}
	\caption{Illustration of the proposed CDUM framework. The light purple block on the left is the offline training part (CPM), and the light green block on the right is the online inference part (FIC).}
	\label{fig:framework}
	\vspace{-3mm}
\end{figure*}

We propose Coarse-to-fine Dynamic Uplift Modeling for Real-time Video Recommendation (CDUM), which includes two modules: Coarse-grained Preference Modeling (CPM) and Fine-grained Interest Capture (FIC). Figure \ref{fig:framework} illustrates the technical details of the proposed model.

\subsection{Coarse-grained Preference Modeling}
In this part, we introduce a module, Coarse-grained Preference Modeling, which utilizes the offline features to model the user's long-term preferences for different treatments.
\subsubsection{Feature Encoder Module.}
\label{sec:cpm_feature}
For each instance $\mathbf{z}_i = (\mathbf{x}_i,\mathbf{t}_i,y_i)$, we encode both $\mathbf{x}_i$ and $\mathbf{t}_i$ into embeddings. For the continuous feature, we first utilize a normalization technique to convert continuous features into sparse features. To be specific, we have:
\begin{equation}
\setlength{\abovedisplayskip}{0.5pt}
\setlength{\belowdisplayskip}{0.5pt}
x_i^{j*} = x_i^j/\mathop{max}\limits_{i}(x_i^j),
\
t_i^{j*} = t_i^j/\mathop{max}\limits_{i}(t_i^j).
\end{equation}
Then, we initialize embedding tables for both processed continuous and sparse features, and capture the embedding representations corresponding to the specific feature value. Specifically, we have:
\begin{equation}
\setlength{\abovedisplayskip}{0.5pt}
\setlength{\belowdisplayskip}{0.5pt}
\mathbf{e}_{xi} = \mathbf{E}_x^{T} \cdot \mathbf{x}_i, \ 
\mathbf{e}_{ti} = \mathbf{E}_t^{T} \cdot \mathbf{t}_i,
\end{equation}
where $\mathbf{E}_x \in \mathbb{R}^{f_x \times d}$ and $\mathbf{E}_t \in \mathbb{R}^{f_t \times d}$ are the created embedding tables, $f_x$ and $f_t$ are the numbers of non-treatment features and treatment features, and $d$ is the embedding size. 
In addition, we can get the corresponding embeddings for each feature, i.e., $\mathbf{e}_{xi}=\{\mathbf{e}_{xi}^0,\mathbf{e}_{xi}^1,...,\mathbf{e}_{xi}^{f_x}\}$ and $\mathbf{e}_{ti}=\{\mathbf{e}_{ti}^0,\mathbf{e}_{ti}^1,...,\mathbf{e}_{ti}^{f_t}\}$

\subsubsection{The Multi-treatment Learning Module}
As we propose to utilize \emph{adjusting the distribution of videos with varying durations} as treatment, the modeling of the problem becomes the paradigm of multi-treatment learning, i.e., learning the preference of users (expressed as responses) for different treatments. First, we encode the non-treatment embeddings of users with a concatenate operation, and use an average pooling operation on the treatment embedding:
\begin{equation}
\setlength{\abovedisplayskip}{0.5pt}
\setlength{\belowdisplayskip}{0.5pt}
    \mathbf{e}_{xi}^{*} = \mathop{Concat}\limits_{j=0}^{f_x}(\mathbf{e}_{xi}^{j}), \mathbf{e}_{ti}^{*} = \sum\limits_{j=0}^{f_t}\mathbf{e}_{ti}^j/(f_t+1),
\end{equation}

Additionally, to fully leverage the treatment information, inspired by multi-interest learning \cite{ckml,mind,comirec}, we further refine treatment embedding into indicator embedding and guidance embedding. We consider these two forms of embeddings as interest generalization under different effects of treatment. For the guidance part, we expect it to serve as a filter and information extractor for user features, expressing the guiding role of the treatment. In contrast, for the indicator part, we expect it to express the indicative role of the treatment, thereby helping the model generalize to learning tasks involving different treatments. Specifically, we first utilize an interest extraction module to refine the guidance and indicator parts from $\mathbf{e}_{ti}$. Take the guidance part as an example, we have:
\begin{equation}
\label{equ:mlp}
\left\{\begin{array}{c}
\begin{aligned}
\mathbf{h}_{ti}^{gui} &=\mathop{ReLU}\left(\mathbf{e}_{ti}^{*}\cdot\mathbf{W}_{gui}^{h}+\mathbf{b}_{gui}^{h}\right), \\
\mathbf{e}_{ti}^{gui} &=\mathop{Sigmoid}\left(\mathbf{h}_{ti}^{gui}\cdot\mathbf{W}_{gui}^{o}+\mathbf{b}_{gui}^{o}\right),
\end{aligned}
\end{array}\right.
\end{equation}
where $\mathbf{W}_{gui}^{h} \in \mathbb{R}^{d \times d_h}$ and $\mathbf{b}_{gui}^{h} \in \mathbb{R}^{1 \times d_h}$ are transformation matrix and bias matrix for the hidden layer, $d_h$ is the dimension of the hidden representation $\mathbf{h}_{ti}^{gui}$. $\mathbf{W}_{gui}^{o} \in \mathbb{R}^{d_h \times d^*}$ and $\mathbf{b}_{gui}^{o} \in \mathbb{R}^{1 \times d^*}$ are transformation matrix and bias matrix for the output layer. $\mathbf{e}_{ti}^{gui}$ is the guidance embedding refined from $\mathbf{e}_{ti}$, and $d^*$ is the dimension of the embedding. While $\mathbf{e}_{ti}^{ind}$ which represents the indicator embedding can be obtained in a similar process.

Furthermore, we first apply operation to $\mathbf{e}_{xi}^{*}$ similar as Equation (\ref{equ:mlp}), and obtain $M$ expert representations (denoted as $\mathbf{f}_m$, $m\in\{1,2,...,M\}$). 
After that, we utilize the guidance embedding to filter and extractor information for user features while leveraging the indicator embedding to enhance the representations of the extracted information, and have:
\begin{equation}
\setlength{\abovedisplayskip}{0.5pt}
\setlength{\belowdisplayskip}{0.5pt}
\left\{\begin{array}{c}
\begin{aligned}
\mathbf{g}_i^{k} &= Softmax(\mathbf{W}_g \cdot \mathbf{e}_{ti}^{gui} + \mathbf{b}_g) \\
\hat{y}_i^{k} &= h^{k}\left(\mathbf{e}_{ti}^{ind}, \sum_{m=1}^{M} {\mathbf{g}_i^{k}(m) \cdot \mathbf{f}_m}\right)
\end{aligned}
\end{array}\right.
\end{equation}
where $\mathbf{W}_g \in \mathbb{R}^{M \times d^*}$ and $\mathbf{b}_g \in \mathbb{R}^{M \times 1}$ are feature transformation matrix and bias matrix, and $\mathbf{g}_i^{k} \in \mathbb{R}^{M \times 1}$ is the attention vector which is used as selector to calculate the weighted sum of all experts. $\mathbf{g}_{i}^{k}(m)$ denotes the $m$-th element of vector $\mathbf{g}_i^{k}$, $h^{k}(\cdot)$ is the tower function. And we use a similar structure with Equation (\ref{equ:mlp}) as the tower function here for simplicity. In addition, $\mathbf{e}_{ti}^{ind}$ is applied as the form of mask before the final output of the tower. $\hat{y}_i^{k}$ denotes the prediction score for the response of user under the $k$-th treatment.

\subsubsection{Training Objective and Prediction.}
\label{sec:cpm_training}
As we have obtained $\hat{y}_i^{k}$ for the response of user under $k$-th treatment, we further use the huber loss \cite{huber_loss} as the training loss function, i.e.,
\begin{equation}
\mathcal{L} = 
\left\{\begin{array}{c}
\begin{aligned}
    &0.5*(\hat{y}_i^{k}-y_i^{k})^2, &if |\hat{y}_i^{k}-y_i^{k}| \leq \delta \\
    &0.5*\delta^2+\delta*(|\hat{y}_i^{k}-y_i^{k}|-\delta), &if |\hat{y}_i^{k}-y_i^{k}| > \delta
\end{aligned}
\end{array}\right.
\end{equation}
where $y_i^{k}$ is the response label under $k$-th treatment. $\delta$ is a constant that can be adjusted dynamically. It is worth noting that in the training stage, since each sample $(\mathbf{x}_i,\mathbf{t}_i,y_i)$ is obtained by recording response only under the action of a specific treatment (e.g, $k$-th treatment), we only use the output $\hat{y}_i^{k}$ of tower corresponding to the treatment for training, while mask operation is applied to the calculation of loss to other towers.

Finally, at the prediction stage, we can infer $\hat{y}_i^{k} (k\in\{0,1,2,...,K\}$) for each user by replacing the feature of treatment $\mathbf{t}_i$  iteratively. Further, we can obtain the long-term preference scores of users for different treatments, and deliver them to the next online part, FIC.

\subsection{Fine-grained Interest Capture}
Although we have captured the long-term preference scores, the targeted decision of the affected group still can not be adjusted dynamically in real time. In the pipeline of online video recommendation, there is rich and varied feature information, such as online real-time contextual features and request-level candidates information. Thus, we innovatively design a module, Fine-grained Interest Capture Module. This module can leverage the online features to model users' real-time interests and dynamically adjust long-term preferences accordingly, enabling real-time dynamic adjustment of the effective user groups for different treatments.

\subsubsection{Module Structue.}
For the selection of features and labels, as mentioned in Section \ref{Online Problem Definition}, we use contextual features and request-level candidates as features, i.e., $\mathbf{s}_i$. At the same time, the ratio of the number of long and short broadcast exposed videos of users under each treatment classification (divided by different durations of videos) is used as the label of corresponding categories. 

We first encode the online features to embeddings. This process is similar to Sectio \ref{sec:cpm_feature}, thus we get the online embeddings $\mathbf{e}_{si} = \{\mathbf{e}_{si}^0,\mathbf{e}_{si}^1,...,\mathbf{e}_{si}^{f_s}\}$. Here $\mathbf{e}_{si}^j \in \mathbb{R}^{1 \times L^j \times d_s}, j \in [0,f_s]$, $L^j$ represents the number of recent videos with which the user has interacted, $d_s$ is the embedding size. Then, we apply an average pooling to the second dimension, and concatenate all the pooled embeddings together. This process can be represented as:
\begin{equation}
\setlength{\abovedisplayskip}{0.5pt}
\setlength{\belowdisplayskip}{0.5pt}
    \mathbf{e}_{si}^* = \mathop{Concat}\limits_{j=0}^{f_s}\left(\sum\limits_{l=1}^{L^j}{\mathbf{e}_{si}^j(l)/(L^j)}\right),
\end{equation}
where $\mathbf{e}_{si}^j(l) \in \mathbb{R}^{1\times1\times d_s}$ denotes the $l$-th interacted item embedding.

Further, we equip $\mathbf{e}_{si}^*$ with a multi-task learning network. Here, the form of the network is similar to MMOE \cite{mmoe}, and for simplify, we denote $\hat{r}_i^k$ ($k \in [1,K]$) as the output of each tower of this part. 

\subsubsection{Training Objective and Prediction}
In the previous section, we have obtained the output $\hat{r}_i^k$ for real-time interest learning of different treatments. As mentioned in Section \ref{sec:cpm_feature}, we design an online request-level label, which is the ratio of the number of long-playing videos to the number of short-playing videos. To be specific, suppose that the number of exposed videos for a user's request is $V$ and we have $K$ categories of treatment. The designation of the label can be formulated as:
\begin{equation}
\setlength{\abovedisplayskip}{0.5pt}
\setlength{\belowdisplayskip}{0.5pt}
\sum\limits_{k=1}^K{V_k} = V, V_k^{long} + V_k^{short} = V_k, r_i^k = \frac{V_k^{long}+1}{V_k^{short}+1},
\end{equation}
where $V_k$ denotes the $k$-th category of video duration (i.e., treatment). $V_k^{long}$ and $V_k^{short}$ represent the numbers of long-playing and short-playing videos, respectively.

Furthermore, for the training, we utilize the huber loss as optimization function, which is similar to the process in Section \ref{sec:cpm_training}. While for the prediction, we deploy this module directly online, and it is invoked with every user request to calculate the real-time interest score $\hat{r}_i^k$.

\subsubsection{Decision Making.}
As we have obtained the offline long-term preference score $\hat{y}_i^{k} (k\in\{0,1,...,K\}$) and online real-time interest score $\hat{r}_i^k (k\in\{1,2,...,K\})$, we can eventually combine them to dynamically adjust the effective user groups for different treatments. Specifically, as shown in Figure \ref{fig:framework}, we define $\hat{r}_i^k*\hat{y}_i^k-\hat{y}_i^0$ as the decision score for the $k$-th treatment. We then define $\xi$ as the threshold for whether the treatment takes effect (i.e., if $\hat{r}_i^k*\hat{y}_i^k-\hat{y}_i^0 > \xi$, the $k$-th treatment is enabled). Thus, the multi-treatment (i.e., adjusting the distribution of videos with varying durations) for each user will be dynamically adjusted to enable or disable in real time.
\section{OFFLINE EVALUATIONS}
\label{experiments}

\subsection{Experimental Setting}
\subsubsection{Dataset Description}

We evaluate our proposed method on both public and real-world industrial datasets. For the public dataset, we use CRITEO \cite{criteo} and LAZADA \cite{descn}. CRITEO is a dataset built by Criteo AI Labs in a real large-scale advertising scenario. It includes nearly 14 million instances, 12 features, a treatment indicator, and two labels (i.e., visit and conversion). We use visit as the target, and randomly split the dataset for the training, validation, and testing with a ratio of 8:1:1. LAZADA is a real voucher distribution business scenario dataset from the e-commerce platform Lazada. It contains 83 features, a treatment indicator, and a label. Due to the operational target strategy, the treatment allocation is selective in the actual production environment, these data containing treatment bias are utilized as training set, and a small number of other users who are not affected by the targeting strategy are used as testing set, and their treatment allocation follows the randomized controlled trial.

The industrial Kuaishou dataset consists of online user data collected 14 days prior to the launch of multiple randomized controlled trial (RCT) experiments with different treatments and non-treatments. The data features include users' inherent attributes (region, activity, etc.), consumption information from the past 7 and 30 days (duration of usage and video views, etc.), and user's commercial attributes (including user commercial value classification). Additionally, we calculated the average app usage time during the experiment as a label. The number of instances in each group (with different treatments) is about 8,500,000. We split the data into training, validation, and test sets in a ratio of 8:1:1.

\subsubsection{Evaluation Metrics}
In our experiments, we adopt three widely used metrics to evaluate uplift ranking performance of different models: normalized Area Under the Uplift Curve (AUUC), normalized area under the qini curve (QINI), and uplift score at first $h$ percentile (LIFT@$h$, $h$ is set as 30). We compute these metrics using a standard python package scikit-uplift\footnote{https://www.uplift-modeling.com/en/latest}.

\subsubsection{Baseline Models}
To demonstrate the effectiveness of our proposed CDUM, we selected a set of currently popular and representative uplift modeling methods, including:  S-Learner \cite{metalearners}, T-Learner \cite{metalearners}, BNN \cite{bnn}, TARNet \cite{shalit2017estimating}, CFRNet \cite{shalit2017estimating}, CEVAE\cite{cevae}, GANITE \cite{ganite}, DragonNet \cite{dragonnet}, FlexTENet \cite{flextenet}, EUEN \cite{euen}, DESCN \cite{descn}, and EFIN \cite{efin}.

For the experiments on the Kuaishou dataset, we utilize some classic multi-task models to replace the CPM model in order to test the effectiveness, including Shared Bottom \cite{sharebottom}, MMOE \cite{mmoe}, PLE \cite{PLE}, MESI \cite{cigf}. Since the dataset contains multiple treatments, the final evaluation metric is the average of the AUUC scores of all treatments, named AUUC$_{avg}$.

\subsubsection{Parameter Settings}
Our proposed CDUM is implemented in TensorFlow \cite{TensorFlow}. We use the Adam optimizer \cite{adam} with a learning rate of 0.001 and set the learning rate reduction factor to 0.6. The number of allowed epochs with no improvement is 2, after which the learning rate will be reduced. In addition, we set the batch size to 4096 and the embedding dimension to 32. The number of expert modules in CPM is 3. The default values of $\delta$ and $\xi$ are 1 and 0, respectively.

\begin{table}[H]
\caption{The overall performance comparison on Lazada and Cretio datasets. Boldface denotes the highest score and underline indicates the results of the best baselines. $\star$ represents significance level $p$-value $<0.05$ of comparing CPM with the best baseline.}
    \centering
    \begin{threeparttable}
	\resizebox{\linewidth}{!}{
    \begin{tabular}{c|ccc|ccc}
    \toprule
    {Dataset}&
    \multicolumn{3}{c}{LAZADA}&\multicolumn{3}{|c}{CRITEO}\cr
    \cmidrule(lr){1-1}
    \cmidrule(lr){2-4} \cmidrule(lr){5-7} 
    {Metrics}&QINI&AUUC&LIFT@30&QINI&AUUC&LIFT@30\cr
    \midrule
    S-Learner &0.0207&0.0029&0.0072&0.0902 & 0.0354 & 0.0298 \cr
    T-Learner &0.0217&0.0031&0.0084&0.0790 & 0.0310 & 0.0277 \cr
    BNN &0.0230&0.0033&0.0073&0.0888 & 0.0349 & 0.0286 \cr
    TARNet &0.0214&0.0030&0.0072&0.0847 & 0.0333 & 0.0288 \cr
    CFRNet &0.0231&0.0033&\underline{0.0087}&0.0901 & 0.0354 & \underline{0.0302} \cr
    CEVAE &0.0158&0.0022&0.0074&0.0867 & 0.0341 & 0.0290 \cr
    GANITE &0.0174&0.0024&0.0078&0.0818 & 0.0322 & 0.0283 \cr
    DragonNet &0.0176&0.0025&0.0077&0.0851 & 0.0335 & 0.0289 \cr
    FlexTENet &0.0185&0.0026&0.0079&\underline{0.0924} & \underline{0.0363} & 0.0298 \cr
    SNet &\underline{0.0238}&\underline{0.0034}&0.0076&0.0843 & 0.0331 & 0.0284 \cr
    EUEN &0.0156&0.0022&0.0073&0.0898 & 0.0353 & 0.0300 \cr
    DESCN &0.0223&0.0031&0.0075&0.0803 & 0.0316 & 0.0274 \cr
    EFIN &0.0141&0.0020&0.0077&0.0859 & 0.0337 & 0.0287 \cr
    \hline    \textbf{CPM}&\textbf{0.0265}$^\star$&\textbf{0.0037}$^\star$&\textbf{0.0087}$^\star$&\textbf{0.0942}$^\star$&\textbf{0.0371}$^\star$&\textbf{0.0310}$^\star$\cr
    \bottomrule
    \end{tabular}}
    \end{threeparttable}
    \label{comparisons_model111}
\end{table}

\subsection{Performance Comparison}
We report the evaluation results of the offline module CPM in our proposed CDUM and the baselines on two offline public datasets in Table 1. From the table, we can see that:
\begin{itemize}
    \item On both two datasets, the performance of S-learner and T-Learner is not bad and even outperforms some baselines using more complex network architectures. This shows that in the scenario of large-scale uplift modeling, complex modeling methods may lead to suboptimal result due to the inability to quickly identify the sensitive features of users.
    \item On the LAZADA dataset, SNet performs better than other baselines overall, but on the CRITEO dataset, the best performing baseline is FlexTENet. In addition, some baselines (e.g., EUEN and DESCN) exhibit significant performance variations when evaluate on the two datasets. This inconsistency highlights that current baseline models are unable to provide reliable performance across different scenarios, lacking strong generalizability.
    \item Our proposed CPM performs best in all metrics of the two datasets, which demonstrates the effectiveness and applicability of our proposed offline module CPM in the task of uplift modeling.
\end{itemize}

To further validate the performance of CPM in a multi-treatment setting and to integrate it with the online module FIC for deployment, we conducted experiments on our Kuaishou dataset. Table \ref{comparisons_model222} shows that CPM still maintains superior performance in the multi-treatment scenario.

\begin{table}[t]
   \setlength{\abovecaptionskip}{0cm}
   \setlength{\belowcaptionskip}{-0.0cm}
\caption{The overall performance comparison on Kuaishou offline dataset. Boldface denotes the highest score and underline indicates the results of the best baselines. $\star$ represents significance level $p$-value $<0.05$ of comparing CPM with the best baseline.}
    \centering
    \begin{threeparttable}
	\resizebox{0.45\linewidth}{!}{
    \begin{tabular}{c|c}
    \toprule
    {Dataset}&Kuaishou \cr
    \cmidrule(lr){1-1}
    \cmidrule(lr){2-2}
    {Metrics} &AUUC$_{avg}$ \cr
    \midrule
    T-Learner &0.00787 \cr
    CausalForestDML &0.00914 \cr
    Shared Bottom &0.00798 \cr
    MMOE &0.00947 \cr
    PLE &0.00998 \cr
    MESI &\underline{0.01053} \cr
    \hline 
    \textbf{CPM}&\textbf{0.01143}$^\star$ \cr
    \bottomrule
    \end{tabular}}
    \end{threeparttable}
    \label{comparisons_model222}
    \vspace{-5mm}
\end{table}

\begin{table}[H]
   \setlength{\abovecaptionskip}{0cm}
   \setlength{\belowcaptionskip}{-0.0cm}
\caption{Performance of different CPM variants. $\star$ represents significance level $p$-value $<0.05$ of comparing CPM with the best variants.}
    \centering
    \begin{threeparttable}
	\resizebox{\linewidth}{!}{
    \begin{tabular}{c|ccc|ccc}
    \toprule
    {Dataset}&
    \multicolumn{3}{c}{LAZADA}&\multicolumn{3}{|c}{CRITEO}\cr
    \cmidrule(lr){1-1}
    \cmidrule(lr){2-4} \cmidrule(lr){5-7} 
    {Metrics}&QINI&AUUC&LIFT@30&QINI&AUUC&LIFT@30\cr
    \midrule
    CPM $w/o$ indicator &0.0218&0.0030&0.0081&0.0922 & 0.0362 & 0.0308 \cr
    CPM $w/o$ guidance &0.0220&0.0031&0.0081&0.0901 & 0.0354 & 0.0304 \cr
    \hline    \textbf{CPM}&\textbf{0.0265}$^\star$&\textbf{0.0037}$^\star$&\textbf{0.0087}$^\star$&\textbf{0.0942}$^\star$&\textbf{0.0371}$^\star$&\textbf{0.0310}$^\star$\cr
    \bottomrule
    \end{tabular}}
    \end{threeparttable}
    \label{ablation}
\end{table}

\subsection{Ablation Study}
In order to evaluate the effectiveness of the key design of the offline module CPM in our proposed method CDUM, we conduct ablation studies. Specifically, we design the following two variants:
\begin{itemize}
    \item CPM w/o indicator: We remove the indicator embedding used to enhance the representation of extracted information.
    \item CPM w/o guidance: We use the initial feature embedding instead of the guidance embedding to generate the weights when aggregating experts in gating.
\end{itemize}

The results are shown in Table \ref{ablation}. It can be seen that removing any design will cause performance degradation. This demonstrates the rationality and effectiveness of our proposed module.

\section{ONLINE APPLICATION}
To further evaluate the online performance of our model in real product scenarios, we apply CDUM by adding a new model in the industrial recommendation service of Kuaishou, which serves billions of users. We randomly select and serve 20\% of users with our model and another 20\% of users with online baseline, then deploy them online from 2024-09-12 to 2024-09-29 with 18 days of online test.

\begin{figure}[H]
	\centering
	\setlength{\belowcaptionskip}{-0.0cm}
	\setlength{\abovecaptionskip}{-0.0cm}
	\includegraphics[width=\linewidth]{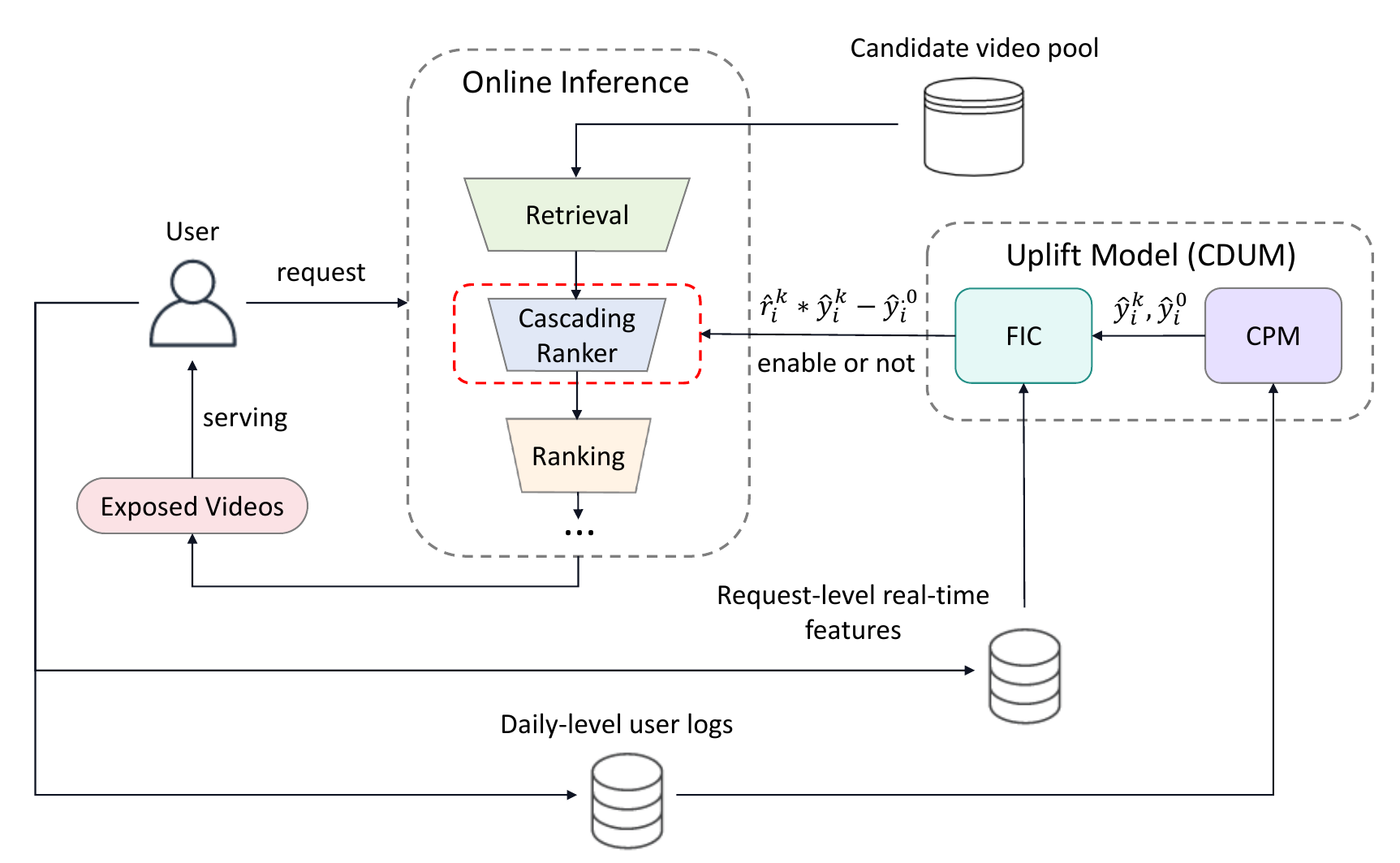}
	\caption{Illustration of the online pipeline.}
	\label{fig:online_pipeline}
	\vspace{-3mm}
\end{figure}

\subsection{System Description}
As shown in Figure \ref{fig:online_pipeline}, in the pipeline of online video recommendation, a series of modules are cascaded together to filter candidate videos and finally expose the selected videos to the user. Specifically, when a user triggers a request, the corresponding user's attributes and contextual features are sent to the online service. And the online pipeline is triggered to filter and select the videos from the candidate pool in a cascading form. At last, the selected videos are exposed to the user, thus the request is over.

\subsection{Scheme Overview}
Our model, as illustrated in Figure \ref{fig:online_pipeline}, is developed at the bucket video duration adjustment part of \emph{Cascading Ranker}. To be specific, when a user sends a request, the retrieval part first retrieves candidate videos from a candidate pool, and sends them to the Cascading Ranker. Therefore, the Cascading Ranker is triggered, and in this stage, CDUM is invoked to inference with the input of user request-level features and user's attributes, etc. In CDUM, the outputs $\hat{y}_i^k, \hat{y}_i^0$ of CPM are updated at a daily level, thus they are directly utilized to be calculated for the decision scores. While the FIC module makes inference based on the incoming features with the user request in real time and outputs $\hat{r}_i^k$. These two parts of CDUM work together to decide the enablement of treatments at the level of the user's request, thus directly acting at the bucket video duration adjustment part of the Cascading Ranker.

\begin{figure*}[t]
\centering
    \setlength{\abovecaptionskip}{0cm}
    \setlength{\belowcaptionskip}{0mm}
	\subfigure[Enter LT]{
        \begin{minipage}[t]{0.48\linewidth}
        \centering
		\label{fig:enter} 
		\includegraphics[width=\linewidth]{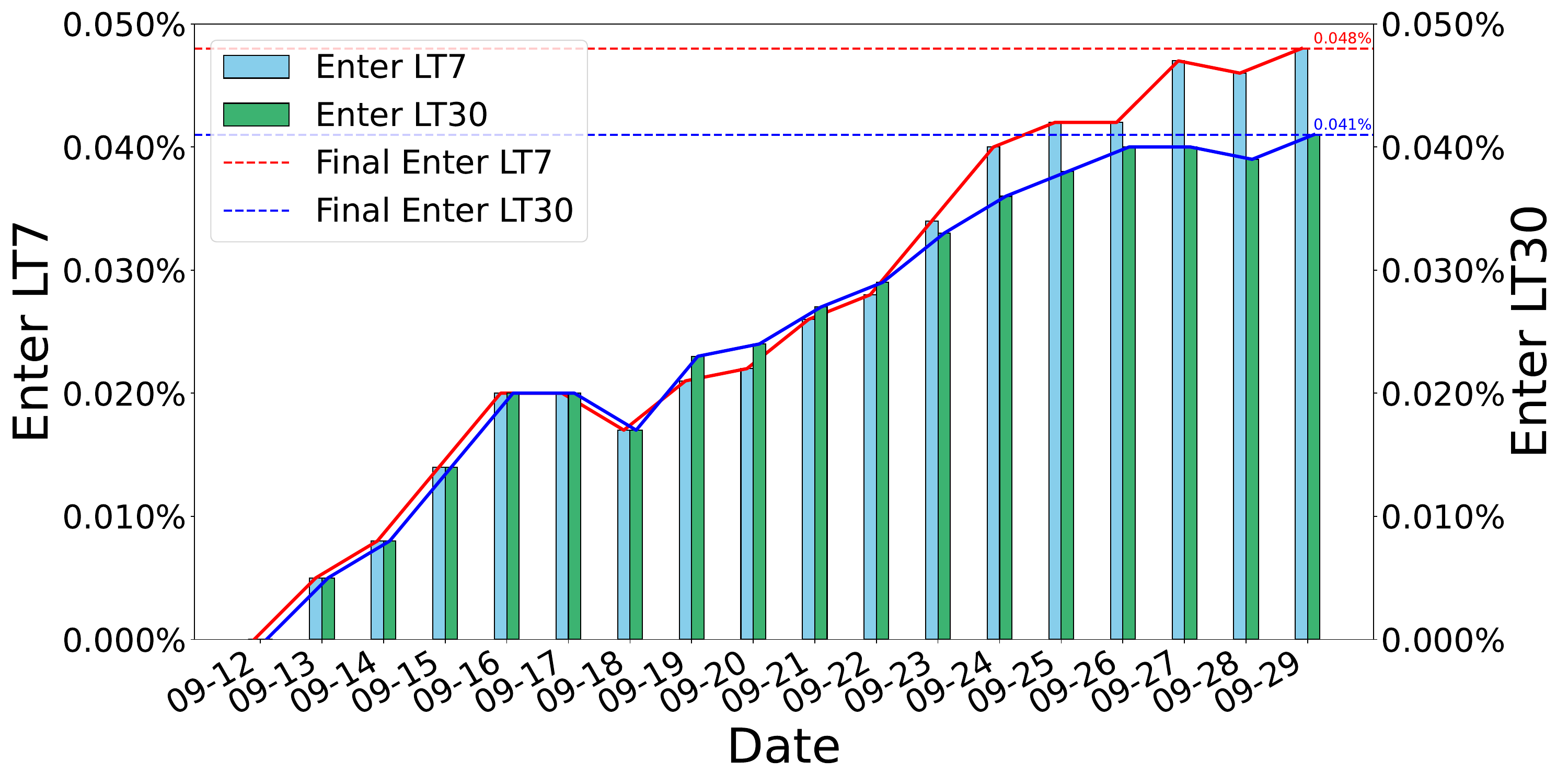}
        \end{minipage}
        }
	\subfigure[Slide LT]{
        \begin{minipage}[t]{0.48\linewidth}
        \centering
		\label{fig:slide} 
		\includegraphics[width=\linewidth]{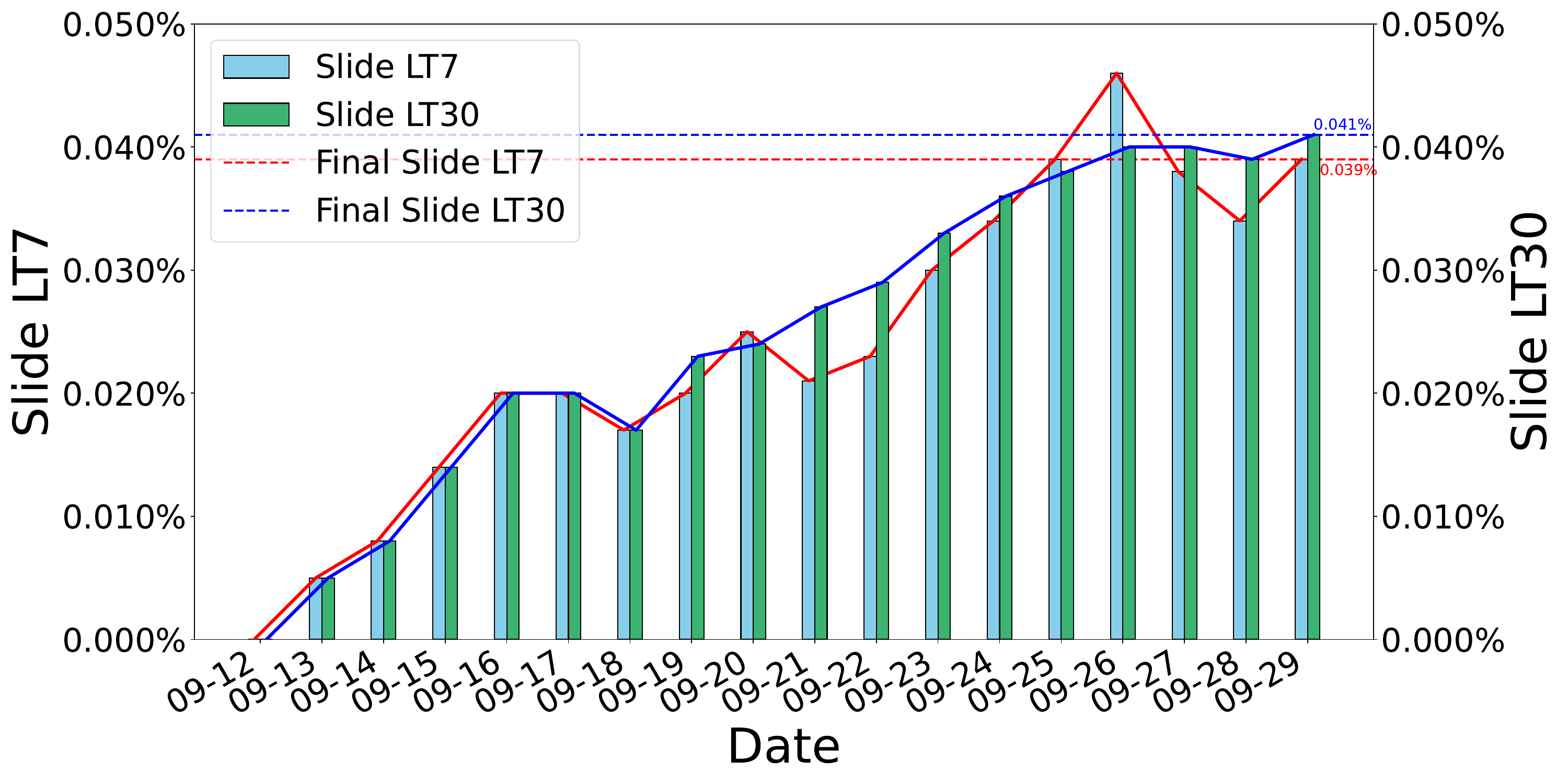}
        \end{minipage}
        }
	\caption{Online A/B test experimental results of Enter LT and Slide LT.}
	\label{fig:online}
\end{figure*}

\subsection{Online Experimental Results} 

\subsubsection{Comparison with Online Baseline.}
We leverage the common consumption metrics (i.e., APP usage time and watch time) and retention metrics (i.e., Next day retention, 7-day retention, Enter LT7/LT30 and Slide LT7/LT30) in the industry to evaluate the online performance. Among them, \textbf{LT7/LT30} is an unbiased estimate of medium- and long-term DAU for experimental evaluation, pointing strictly to user experience. They are key metrics that indicate user retention benefits online, and can be formulated as:
\begin{equation}
\left\{\begin{array}{c}
\begin{aligned}
    \text{Enter LT7}  &= \frac{\sum_{i=max(T-6,T_0)}^{T}DAU_i}{\text{Total number of active users from } T_0 \text{ to } T}\\
    \text{Slide LT7}  &= \frac{\sum_{i=max(T-6,T_0)}^{T}DAU_i}{WAU}
\end{aligned}
\end{array}\right.
\end{equation}
where $T$ stands for the date the experiment data is recorded, $T_0$ denotes the date that the experiment is started. $DAU_i$ represents the number of daily active users at the $i$-th day. $WAU$ represents the number of weekly active users calculated from $max(T-6,T_0)$ to $T$. The LT30 is calculated in a similar way.
And as Kuaishou serves hundreds of millions of users every day, it is known that a slightly higher LT7/LT30 (↑) at \emph{\textbf{0.01\%}-level} is regarded significant.

Figure \ref{fig:online} shows the performance of LT metrics of our model compared to the online baseline during the test period. We can find that our model achieves significant improvements of \textbf{0.048\%}/\textbf{0.041\%} and \textbf{0.041\%}/\textbf{0.039\%} with respect to Enter LT7/LT30 and Silde LT7/LT30 respectively, which verifies the effectiveness of our proposed method. Besides, as shown in Table \ref{online_ab111}, our model shows significant improvement over baseline in other retention and consumption metrics, again demonstrating the superiority of CDUM. Last but not least, our model is now deployed online with full users at Kuaishou, serving for hundreds of millions of users every day.

\begin{table}[t]
\caption{The performance of Online A/B Testing at Kuaishou platform. Boldface represents significance level $p$-value $<0.05$ of comparing CDUM with the online baseline.}
    \centering
    \begin{threeparttable}
	\resizebox{0.8\linewidth}{!}{
    \begin{tabular}{c|cc}
    \toprule
    {Dataset}&
    \multicolumn{2}{c}{Kuaishou} \cr
    \cmidrule(lr){1-1}
    \cmidrule(lr){2-3}
    {Metrics} &Online Baseline&\textbf{CDUM}\cr
    \midrule
    Next day retention &-&\textbf{+0.034\%}\cr
    7-day retention &-&\textbf{+0.038\%}\cr
    APP usage time &-&\textbf{+0.135\%}\cr
    APP usage time (per capita) &-&\textbf{+0.131\%}\cr
    watch time &-&\textbf{+0.188\%}\cr
    \bottomrule
    \end{tabular}}
    \end{threeparttable}
    \label{online_ab111}
\end{table}

\subsubsection{Ablation Test of CDUM.}

Furthermore, to demonstrate the effectiveness of the online module FIC, we conduct an ablation experiment for our proposed CDUM. To be specific, we remove the decision-making part of online adjustment, and rely solely on $\hat{y}_i^k-\hat{y}_i^0 > \xi$ ($k\in[1,K]$) to determine whether the $k$-th treatment is enabled or not.

As illustrated in Table \ref{online_ab222}, CDUM significantly outperforms its variant CDUM $w/o$ FIC on consumption metrics, proving the indispensability of FIC.

\begin{table}[t]
\caption{The ablation result of Online A/B Testing at Kuaishou platform. Boldface represents significance level $p$-value $<0.05$ of comparing CDUM with the variant.}
    \centering
    \begin{threeparttable}
	\resizebox{0.8\linewidth}{!}{
    \begin{tabular}{c|cc}
    \toprule
    {Dataset}&
    \multicolumn{2}{c}{Kuaishou} \cr
    \cmidrule(lr){1-1}
    \cmidrule(lr){2-3}
    {Metrics} &CDUM $w/o$ FIC&\textbf{CDUM}\cr
    \midrule
    APP usage time &-&\textbf{+0.078\%}\cr
    APP usage time (per capita) &-&\textbf{+0.094\%}\cr
    watch time &-&\textbf{+0.129\%}\cr
    \bottomrule
    \end{tabular}}
    \end{threeparttable}
    \label{online_ab222}
\end{table}
\section{CONCLUSIONS AND FUTURE WORK}
\label{future works}
In this paper, we explore uplift modeling in the context of video recommendation. To appropriately design treatments that capture user's real-time interests while watching videos, we propose Coarse-to-fine Dynamic Uplift Modeling (CDUM) for real-time video recommendation. This model leverages a Coarse-grained Preference Modeling (CPM) module to learn users' long-term preferences using their offline daily-level features, and then employs a Fine-grained Interest Capture (FIC) module for online inference based on real-time features, thereby capturing users' immediate interests. Extensive offline experiments on public and industrial datasets as well as online A/B test on the Kuaishou platform have demonstrated the effectiveness and superiority of our method.

For future work, we plan to explore the potential benefits of enhancing modeling at other stages of the recommendation pipeline, such as ranking, while further designing and improving the model structure and the utilization of features.


\bibliographystyle{ACM-Reference-Format}
\bibliography{sample-base}

\appendix

\end{document}